\documentclass[12pt]{article}
\usepackage{epsfig}

\def\be{\begin{equation}}
\def\ee{\end{equation}}
\def\ba{\begin{array}{c}}
\def\ea{\end{array}}

\def\ben{$$}
\def\een{$$}
\newcommand{\bbr}{\br\!\br}

\newcommand{\pbr}{\prec\!}
\newcommand{\pkt}{\!\succ}
\newcommand{\kt}{\rangle}
\newcommand{\br}{\langle}
\begin{document}

\titlepage

 \begin{center}{\Large \bf

Quantum knots

 }\end{center}

\vspace{5mm}

 \begin{center}
Miloslav Znojil

\vspace{3mm}

Nuclear Physics Institute ASCR, 250 68 \v{R}e\v{z}, Czech
Republic\footnote{ e-mail: znojil@ujf.cas.cz }

\end{center}

\vspace{5mm}

\section*{Abstract}

%
%

We construct an exactly solvable example of Sturmian bound states
which exist in the absence of any confining potential. Their
origin is topological -- these states are found to live on certain
``knotted" contours ${\cal C}^{(N)}$ of complexified coordinates.

 \vspace{9mm}

\noindent
 PACS 03.65.Ge, 11.10.Kk, 11.30.Na, 12.90.+b



\newpage

\section{Introduction \label{s1} }

A {physical} framework and motivation of our forthcoming
considerations lies partially (though not only) in the standard
radial Schr\"{o}dinger equation
 \be
  -\frac{d^2}{d\xi ^2}\,\psi (\xi)+ \frac{\ell(\ell+1)}{\xi^2}
  \,\psi (\xi)+ \lambda\,V(\xi^2)
  \,\psi (\xi)= E \,\psi (\xi)\,
   \label{SEnotfree}
 \ee
where the radial coordinate $\xi$ runs over the half-axis
${I\!\!\!R}^+$ and where the standard Dirichlet boundary
conditions are usually imposed at $\xi = \infty$ and in the origin
(at $\xi=0$, with the well known exceptions for strongly singular
$V(\xi^2)$ \cite{pra}). In phenomenological setting, such an
ordinary differential equation is usually obtained from a
``realistic", spherically symmetric $D-$dimensional
single-particle Hamiltonian $\tilde{H}=-\triangle
+\lambda\,V(|\vec{x}|^2)$ acting in the most common representation
${\it I\!\!\!L}^2({\it I\!\!\!R}^D)$ of the Hilbert space of bound
states. With $\ell=(D-3)/2+ m$ in the $m-$th partial wave, one has
to distinguish between $D=1$ and $D>1$ \cite{BG}. At $D=1$ the
situation is exceptional and a due care is needed when one tries
to work, purely formally, with $m=0$ (for the even-parity states)
and $m=1$ (for the odd-parity states).  At all the higher
dimensions $D \geq 2$, the correspondence between $\tilde{H}$ and
eq.~(\ref{SEnotfree}) is more standard and the sequence of the
angular-momentum indices becomes infinite, $m=0, 1, \ldots$. In
practice, one usually works with a fixed strength $\lambda=1$ of
the interaction and studies the spectrum of the bound-state
energies $E_n$, $n = 0, 1, \ldots$. Alternatively, one can choose
and fix the energy (say, $E=1$) and compute the related
eigencouplings $\lambda_n$, $n = 0, 1, \ldots$ which correspond to
the normalizable solutions $\psi (\xi)$ called Sturmians (this
will also be our choice in what follows).

An immediate {mathematical } inspiration of our present note can
be traced back to the 1993 paper by Buslaev and Grecchi \cite{BG}
who {\em complexified}, purely formally, the variable $\xi$ in
eq.~(\ref{SEnotfree}) (for more details cf. section \ref{prva}
below). From the historical point of view it proved a bit
unfortunate that the Buslaev's and Grecchi's considerations did
not attract too much attention. It took further five years before
Bender, Milton and Boettcher \cite{BB} returned to the subject,
revealed and emphasized its formal appeal and persuaded many
physicists about many phenomenological potentialities hidden in
similar models. In this setting, our present brief note can be
read as a part and a continuation of the deeper analysis of the
{\em redefined} models (\ref{SEnotfree}) using {\em complex} $\xi$
which, strictly speaking, shouldn't be called a ``particle
coordinate" anymore \cite{Batal}.

We shall study the maximally simplified, analytically solvable
version of the Schr\"{o}dinger differential equation without any
interaction,
 \be
  -\frac{d^2}{d\xi^2}\,\psi (\xi)+ \frac{\ell(\ell+1)}{\xi^2}
  \,\psi (\xi)= E \,\psi (\xi)\,.
   \label{SEfree}
    \label{SEto}
    \label{SEr}
 \ee
For compensation, the complexification of coordinates will be
assumed more sophisticated than usual. In the formal definition of
$\xi \in {\cal C}^{(N)}$, the complex contours ${\cal C}^{(N)}$
will be specified as highly unusual and topologically nontrivial
(cf. section \ref{druzka} below). As a consequence, we shall be
able to obtain bound states by imposing the corresponding more or
less standard complexified asymptotic boundary conditions (cf. a
broader context outlined in refs. \cite{BB} and/or
\cite{tob,tobscatt,tob2}). The simplicity of the dynamics encoded
in eq.~(\ref{SEr}) will enable us to construct our bound-state
solutions in closed analytic form (cf. section \ref{soudruzka}). A
more detailed discussion concerning the interpretation and
perspectives of applicability of our $\psi (\xi) \in {\it
I\!\!\!L}^2({\cal C}^{(N)})$ will be added in section \ref{drzka}
and in a brief summary.

Marginally, let us note that sometimes, one could need a slight
extension of the scope of our model (\ref{SEr}) beyond its purely
kinematical version. This can be easily achieved by an addition of
a trivial potential $V(r) = \gamma/r^2$ and by the subsequent
redefinition of the effective $\ell$ in (\ref{SEr}),
 \be
 \ell(\ell+1) = \gamma + \left (m+\frac{D-3}{2}
 \right )\,\left (m+\frac{D-1}{2}
 \right )\,,\ \ \ \ \ \ \
 m=0, 1, \ldots\,.
 \ee
In this way one can treat $\ell=\ell(\gamma)$ as a continuous, not
necessarily just a (half)integer real parameter.

\section{The Buslaev's and Grecchi's model as a guide  \label{prva} }

\subsection{Isospectral Hamiltonians}

Let us briefly return to the Buslaev's and Grecchi's paper
\cite{BG} where a constant shift $\epsilon> 0$ has been used to
define the following straight line of ``unmeasurable" complexified
coordinates,
 \be
  {\cal C}^{(BG)} = \{\xi =
  x-{\rm i} \epsilon\,|\, \epsilon> 0\,,
  \label{conto}
  \,x \in I\!\!R\}\,.
 \ee
A very specific anharmonic-oscillator potential has further been
chosen as  acting along ${\cal C}^{(BG)}$. Under the most common
Dirichlet asymptotic boundary conditions one reveals that with
$\psi [\xi(\pm \infty)]=0$ we have $\phi^{(BG)}(x)\,\equiv\,\psi
[\xi(x)] \in {\it I\!\!\!L}^2({\cal C}^{(BG)})$ obtainable from
the differential equation
 \be
  \left [
  -\frac{d^2}{dx^2}+ \frac{\ell(\ell+1)}{[\xi(x)]^2}
  + \lambda\,V^{(BG)}\{[\xi(x)]^2\}- E^{(BG)}
  \right ]
  \,\phi^{(BG)} (x) =0\,.
   \label{SEnotfreeBG}
 \ee
This is a non-Schr\"{o}dinger, non-selfadjoint eigenvalue problem
with ${\cal PT}-$symmetry defined in terms of the spatial
reflection ${\cal P}$ and temporal reflection ${\cal T}$ and
exhibited by the Hamiltonian $H^{(BG)}$ \cite{BG}.

After a ``naive"  choice of the Hilbert space ${\cal
H}^{(original)}\,\equiv\, {\it I\!\!\!L}^2({\cal C}^{(BG)})$ the
Buslaev's and Grecchi's Hamiltonian $H^{(BG)}$ proves manifestly
non-Hermitian (and, hence, apparently ``unphysical"). Fortunately,
one of the main results of ref. \cite{BG} tells us that $H^{(BG)}$
proves isospectral to another operator
 \be
  h^{(BG)}
 =
 \,\Omega\,H^{(BG)}\,\Omega^{-1}\,
 \label{mapppp}
 \ee
which happens to be self-adjoint and, hence, {\em physical}. This
observation settled the questions of physics beyond BG model and
re-established the correct probabilistic interpretation of all the
observables in the system in question.

Several papers (cf., e.g., \cite{ptho} or \cite{Jones})
re-analysed the Buslaev's and Grecchi's conclusions recently. This
partially motivated also our forthcoming considerations. One of
our reasons was that for the model $H^{(BG)}$ it was trivial to
guarantee, by construction, that the Hamiltonian $h^{(BG)}$
becomes {self-adjoint in its own} Hilbert space ${\cal
H}^{(physical)}$. The challenge of a search for some other simple
models was imminent.

\subsection{General formalism and an amended Dirac's notation}

On a formal level needed in our forthcoming considerations one
should refer to the review paper \cite{Geyer} where the authors
emphasized that $ h^{(BG)}$ in (\ref{mapppp}) can be self-adjoint
(in the Dirac's transposition-plus-complex-conjugation sense,
i.e.,  $ h^{(BG)}= \left (h^{(BG)}\right )^\dagger$) only if $
H^{(BG)}$ is quasi-Hermitian (i.e., only if $  \left
(H^{(BG)}\right )^\dagger=\Theta\,H^{(BG)}\,\Theta^{-1}$ where, in
our present notation, $\Theta=\Omega^\dagger\Omega$). From such a
point of view, the Buslaev's and Grecchi's original choice of
their very specific anharmonic-oscillator model can be interpreted
as a ``mixed blessing". On the negative side, the narrow-minded
results of ref.~\cite{BG} did not prove too inspiring. In fact,
they looked so exceptional that the physics community accepted
them as a mere mathematical curiosity. On the positive side, the
tractability of the model seems to have opened new perspectives.

The point is that in principle, the lower-case ``correct"
Hamiltonian operator can be interpreted as acting in {\em
another}, different Hilbert space ${\cal H}^{(physical)}$.  Thus,
the Hamiltonian $h^{(BG)}$ can be, in general, {\bf very}
different from its original upper-case representation introduced
as acting in a ``tentative", unitarily non-equivalent Hilbert
space ${\cal H}^{(original)}$. Exceedingly complicated versions of
the ``physical" $ h^{(BG)}$ may be encountered in some realistic
models, e.g., in nuclear physics \cite{Geyer}.

In the standard Dirac's notation all the elements $|\Psi \kt$ of
the original vector space and of its dual $\left({\cal
H}^{(original)}\right)^\dagger $ may be treated and denoted as the
usual kets $|\Psi \kt$ and bras $\br \Psi|$, respectively. After
the change of the spaces ${\cal
H}^{(original)}\,\longrightarrow\,{\cal H}^{(physical)}$ it is
necessary to keep the trace of the changes in order to avoid the
possible ambiguity of the notation. Thus, the elements of the
physical Hilbert space will be denoted here by the specific, curly
ket symbols  $|\Psi \pkt\ \in {\cal H}^{(physical)}$ while in the
dual space of linear functionals~\cite{Messiah} we shall write
$\pbr \Psi|\ \in \left ({\cal H}^{(physical)} \right )^\dagger$.

Puzzling as it may seem at the first sight, our emphasis on the
difference between the spaces of kets  $|\Psi \pkt\ \in {\cal
H}^{(physical)} $ and  $|\Psi \kt  \in {\cal H}^{(original)}$ did
in fact play a key role in some misunderstandings which appeared
in the current literature \cite{tdep}. Paradoxically, the most
natural and transparent resolution of the whole puzzle is
virtually trivial. According to our recent proposal \cite{whichop}
one can simply add an auxiliary,  {\em third} Hilbert space ${\cal
H}^{(third)}$ exhibiting the following properties:

\begin{itemize}

\item as a vector space without inner product, the set ${\cal
H}^{(third)}$  {\em coincides} with ${\cal H}^{(original)}$, i.e.,
we may write $|\Psi \kt \in {\cal H}^{(physical)}$ as well;

\item the spaces of duals (or, if you wish, linear functionals)
are {\em different}, i.e., $ \left ({\cal H}^{(original)}
\right)^\dagger := {\cal T}^{(original)}{\cal H}^{(original)} \,
\neq \, \left ({\cal H}^{(third)} \right)^\ddagger := {\cal
T}^{(third)}{\cal H}^{(third)} $;

\item for the auxiliary, innovated conjugation $^\ddagger$ and
functionals $\bbr \Psi | \in \left ({\cal H}^{(third)}
\right)^\ddagger $ we have to postulate the defining relation
 \be
 \bbr \Psi |
 :=
 \br \Psi |\Theta\,\equiv \,
 \left ( ^{\mbox{}}|\Psi \kt^{}
 \right )^\ddagger\,\neq \,
 \left ( ^{\mbox{}}|\Psi \kt^{}
 \right )^\dagger\,
 ,\ \ \ \ \ \ \ \
 \Theta = \Omega^\dagger\,\Omega\,.
 \label{metric}
 \ee

\end{itemize}

 \noindent
As long as the simultaneous use of both the conjugations would
almost certainly lead to dangerous confusions, we shall always
employ just the Dirac's transposition-plus-complex-conjugation one
here. This means that $ {\cal T}^{(original)}$ or ${\cal
T}^{(physical)}$ will be both characterized by the same
single-cross superscripts $^\dagger$ and by the usual bra-ket
correspondence. In contrast, the double cross $^\ddagger$ will not
be used at all. Thus, we shall {\em always} treat the space $
\left ({\cal H}^{(third)} \right)^\ddagger $ and its double-bra
elements $\bbr \Psi | \neq \br \Psi |$ as mere abbreviations.

All these conventions are summarized in Table~\ref{jednat}. They
immediately imply that $ \pbr \psi|\psi' \pkt = \bbr \psi|\psi'
\kt$ so that the spaces ${\cal H}^{(physical)}$ and ${\cal
H}^{(third)}$ are, by construction, unitarily equivalent. Any one
of them may be employed as physical, therefore. Of course, the
same language and physical interpretation is applicable not only
to the BG model but also to all the models with real spectra
(including the quantum knots to be described below) which may only
look non-Hermitian due to the naive initial choice of the ``wrong"
inner product in ${\cal H}^{(original)}$.

\begin{table}[h]
\caption{The triplet of Hilbert spaces in quantum mechanics}
\label{jednat}
\begin{center}
\begin{tabular}{||c|c|c|c|c||}
\hline \hline
 {\rm Hilbert  space} &{\rm element}&{\rm \bf dual} & {\rm \it
  inner product}&  {\rm  Hamiltonian}
 \\
 \hline
  \hline
 ${\cal H}^{(original)}$ & $|\psi\kt$ & $\br \psi| =
 \left (|\psi\kt
 \right )^\dagger$ &
 $ \br \psi|\psi' \kt$ & $ H^{(BG)} \neq \left ( H^{(BG)}
 \right )^\dagger$\\
  \hline
 ${\cal H}^{(physical)}$& $|\psi\!\pkt\ \equiv\,\Omega|\psi\kt$ &
  $\pbr \psi|=\br
 \psi|\Omega^\dagger$ & $ \pbr \psi|\psi' \pkt$ & $ h^{(BG)}
 =\left ( h^{(BG)}
 \right )^\dagger $
   \\
 \hline
 ${\cal H}^{(third)}$& $|\psi\kt$ &
  $\br\!\br \psi|\,\equiv\ \pbr
 \psi|\Omega$ & $ \bbr \psi|\psi' \kt$ &
 $ H^{(BG)}= \left ( H^{(BG)}
 \right )^\ddagger$
 \\ \hline \hline
\end{tabular}
\end{center}
\end{table}


%
\begin{figure}[h]                     
\begin{center}                         
\epsfig{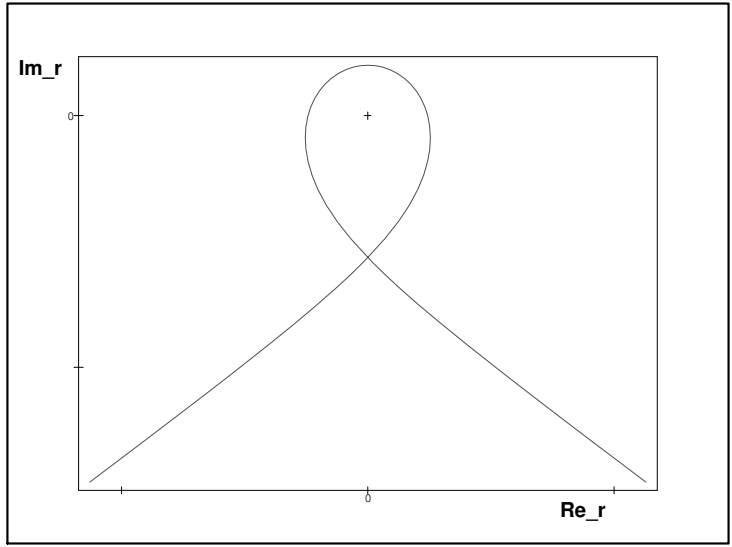}
\end{center}                         
\vspace{-2mm} \caption{Sample of the curve ${\cal C}^{(N)}$ with
$N=1$.
 \label{fione}}
\end{figure}

%
%
\begin{figure}[h]                     
\begin{center}                         
\epsfig{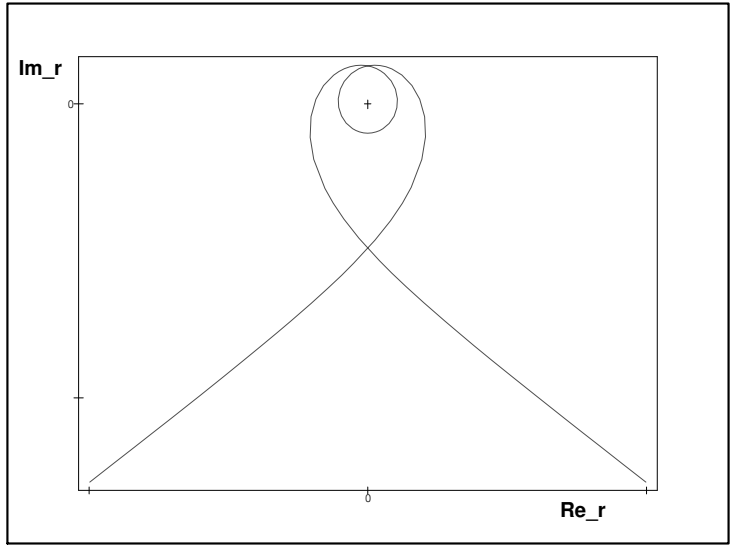}
\end{center}                         
\vspace{-2mm} \caption{Sample of the curve ${\cal C}^{(N)}$ with
$N=2$.
 \label{fitwo}}
\end{figure}

\section{Integration paths
${\cal C}^{(N)}$ \label{druzka} }

Let us now return to our ordinary linear differential
eq.~(\ref{SEr}) of second order, the general solution of which can
always be expressed as a superposition of some of its two linearly
independent components. In particular, in a small complex vicinity
of the origin we may write
 \be
 \psi(\xi) = c_+\,\psi^{(+)}(\xi) +c_-\,\psi^{(-)}(\xi)\,
 \label{ansama}
 \ee
where
 \be
 \psi^{(+)}(\xi) = \xi^{\ell+1} + corrections\,,\ \ \ \
 \psi^{(-)}(r) = \xi^{-\ell} + corrections\,,\ \ \ \
 \ \ \
 \ \ \ |\xi| \ll 1\,.
 \label{mala}
 \ee
In the asymptotic domain we shall prefer another option with
$\kappa=\sqrt{E}$ in
 \be
 \psi(\xi) = c_1\,\psi^{(1)}(\xi) +c_2\,\psi^{(2)}(\xi)\,
 \label{ansala}
 \ee
where
 \be
 \psi^{(1,2)}(\xi) =
 \exp \left ({\pm {\rm i}\,\kappa\,\xi}
 + corrections\right )+ corrections\,,\ \ \ \
 \ \ \ |\xi| \gg 1\,.
 \label{asymptotics}
 \ee
In between these two extremes, our differential eq.~(\ref{SEfree})
is smooth and analytic so that we may expect that all its
solutions are {locally} analytic.

In the vicinity of the origin $\xi=0$ our centrifugal pole with
its real parameter $\ell$ dominates our eq.~(\ref{SEfree}). Once
$\ell$ is assumed irrational, both the components of our wave
functions (as well as their arbitrary superpositions) would
behave, {\em globally}, as multivalued analytic functions defined
on a certain multisheeted Riemann surface ${\cal R}$. In the other
words, our wave functions would possess a logarithmic branch point
in the origin, i.e. a branch point with an infinite number of
Riemann sheets connected at this point \cite{BpT}.

Separately, one should study the simplified models with the
rational $\ell$s which correspond to the presence of an algebraic
branch point at $\xi=0$. A finite number of sheets \cite{BpT}
would be connected there. In the simplest possible scenario of
such a type we may take $\ell(\ell+1)=0$ with either $\ell=-1$ or
$\ell=0$. In such a setting, eq.~(\ref{ansama}) just separates
$\psi(\xi)$ into its even and odd parts so that the Riemann
surface itself remains trivial, ${\cal R}\, \equiv \,l\!\!\!C$.

In the generic case of a multisheeted  ${\cal R}$ we intend to
show that the asymptotically free form of our differential
eq.~(\ref{SEfree}) with the independent solutions
(\ref{asymptotics}) {\em can} generate bound states. One must
exclude, of course, the contours running, asymptotically, along
the real line of $\xi $ since, in such a case, {\em both} our
independent solutions $\psi^{(1,2)}(\xi)$ remain oscillatory and
non-localizable. The same exclusion applies to the parallel,
horizontal lines ${\cal C}^{(BG)}$ in the complex plane of $\xi$.
In the search for bound states, both the ``initial" and ``final"
asymptotic branches of our integration paths ${\cal C}^{(N)}$ must
have the specific straight-line form $\xi=\pm |s|\,e^{{\rm
i}\varphi}$ with a non-integer ratio $\varphi/\pi$. Thus, we may
divide the asymptotic part of the complex Riemann surface of
$\xi\in {\cal R}$ into the sequence of asymptotic sectors
 \begin{equation}
 {\cal S}_0 = \{\xi=-{\rm
i}\,\varrho\,e^{{\rm i}\,\varphi}\,|\,\varrho \gg 1\,,\ \varphi
\in (-\pi/2,\pi/2)\} \label{hola}\,.
 \ee
\be
   {\cal S}_{\pm k} = \{\xi=-{\rm i}\,e^{\pm {\rm
i}\,k\,\pi}\,\varrho\,e^{{\rm i}\,\varphi}\,|\,\varrho \gg 1\,,\
\varphi \in (-\pi/2,\pi/2)\},\ \ \ \ k = 1, 2, \ldots\,.
\label{holas}
 \ee
We are now prepared to define the integration contours ${\cal
C}^{(N)}$. For the sake of convenience we shall set all their
``left" asymptotic branches ${\cal C}^{(left)}$ in the same sector
${\cal S}_{0}$ and specify $\xi=\left (s+s_0\right )\,(1+{\rm
i}\varepsilon)$ where $s \in (-\infty, -s_0)$, $\varepsilon>0$ and
$s_0>0$. The subsequent middle part of ${\cal C}^{(N)}$ must make
$N$ counterclockwise rotations around the origin inside ${\cal R}$
while $s \in (-s_0,s_0)$. Finally, the ``outcoming" or ``right"
asymptotic branch of our integration contour ${\cal C}^{(N)}$ with
$s \in (s_0, \infty)$ must lie in another sector ${\cal S}_{2N}$
of ${\cal R}$, i.e., in the Riemann sheet where the requirement of
${\cal PT}-$symmetry \cite{tobscatt} forces us to set $\xi=\left (
s-s_0\right )\,(1-{\rm i}\varepsilon)$.

\begin{figure}[h]                     
\begin{center}                         
\epsfig{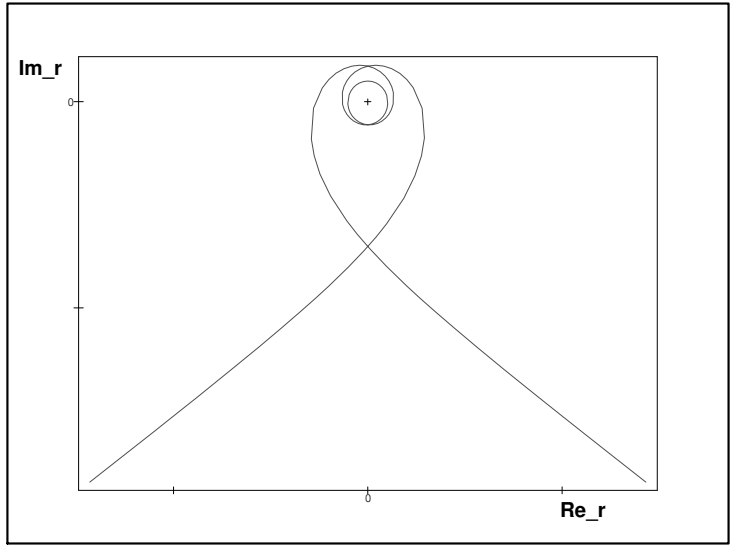}
\end{center}                         
\vspace{-2mm} \caption{Sample of the curve ${\cal C}^{(N)}$ with
$N=3$.
 \label{fithree}}
\end{figure}

\section{Bound states along nontrivial paths \label{soudruzka}}

From now on we shall assume that the integration contour ${\cal
C}^{(N)}$ is fixed and that the variability of $\xi$ is confined
to it. In this spirit we also adapt our notation writing $\xi =
\xi(s)\,\equiv\, r \in {\cal C}^{(N)}$. Our illustrative Figures
\ref{fione} -- \ref{fithree} sample the choice of $N=1$, $N=2$ and
$N=3$, respectively.

It remains for us to impose the asymptotic boundary conditions
requiring that our wave functions vanish at $s \to \pm \infty$. As
long as our integration path ${\cal C}^{(N)}$ performs $N$
counterclockwise rotations around the origin, this form of the
asymptotic boundary conditions will already guarantee the
normalizability of our bound-state wave functions $\psi (r)\in
{\it I\!\!\!L}^2\left ({\cal C}^{(N)} \right ) $ (cf. the similar
situation encountered in the models with confining
potentials~\cite{BG,tob,Alvarez}).

In our bound-state problem (\ref{SEto}) with the (by assumption,
real) $E=\kappa^2$  we may set $z=\kappa r$ and
$\psi(r)=\sqrt{z}\,\varphi(z)$. This reduces eq.~(\ref{SEto}) to
the Bessel differential equation with the pair of the two well
known independent special-function (say,
Hankel-function~\cite{Ryzhik}) solutions which may be inserted in
our ansatz
 \be
 \psi(r) = c_1\,\sqrt{r}\,H^{(1)}_\nu(\kappa\,r)
 +c_2\,\sqrt{r}\,H^{(2)}_\nu(\kappa\,r)\,,\ \ \ \  \nu =
 \ell+1/2\,.
 \label{ansat}
 \ee
At $|{\rm arg}\,z| < \pi$ and ${\rm Re}\,\nu > -1/2$, the
asymptotics of its components are given by the respective formulae
8.451.3 and 8.451.4 of ref.~\cite{Ryzhik},
 \ben
 \sqrt{\frac{\pi z}{2}}\,H^{(1)}_\nu(z)=
 \exp\left [{\rm i}\left (z-\frac{\pi(2\nu+1)}{4}
 \right )\right ]\,\left (1-\frac{\nu^2-1/4}{2{\rm i} z} + \ldots
 \right )
 \,,
 \een
 \ben
 \sqrt{\frac{\pi z}{2}}\,H^{(2)}_\nu(z)=
 \exp\left [-{\rm i}\left (z-\frac{\pi(2\nu+1)}{4}
 \right )\right ]\,\left (1+\frac{\nu^2-1/4}{2{\rm i} z} + \ldots
 \right )
 \,.
 \een
This implies that inside the even-subscripted sectors ${\cal
S}_{2k}$ our ansatz (\ref{ansat}) combines the asymptotically
growing (and, hence, unphysical) component $H^{(1)}_\nu(z)$ with
the asymptotically vanishing and normalizable, physical component
$H^{(2)}_\nu(z)$. {\it Vice versa}, in all the odd-subscripted
sectors ${\cal S}_{2k+1}$ we would have to eliminate, in
principle, the asymptotically growing $H^{(2)}_\nu(z)$ and to keep
the asymptotically vanishing $H^{(1)}_\nu(z)$.

We may start our discussion of the existence of the localized
bound states from the straight-line contour ${\cal C}={\cal
C}^{(BG)} = {\cal C}^{(0)}$ which is all contained in the zeroth
sector ${\cal S}_0$. This immediately implies that with ${\rm
Im}\,r \ll -1$, the asymptotically vanishing solution
 $
 \psi^{(1)}(r) =
 \sqrt{ r}\,H^{(2)}_\nu(\kappa\,r)$
remains unconstrained at all the real $\kappa$. Obviously, the
spectrum remains non-empty and bounded from below. This means that
the low-lying states remain stable with respect to a random
perturbation. A less usual feature of such a model is that its
energies densely cover all the real half-line $I\!\!R^+$. This
feature is fairly interesting {\em per se}, although a more
detailed analysis of its possible {\em physical} consequences lies
already beyond the scope of our present brief note.

Our eigenvalue problem becomes not too much more complicated when
we turn attention to the spiral- or knot-shaped integration
contours ${\cal C}^{(N)}$ with~$N> 0$. In such a case,
fortunately, the exact solvability of our differential equation
enables us to re-write ansatz (\ref{ansala}) in its fully explicit
form which remains analytic on all our Riemann surface ${\cal R}$.
Once we choose our ``left" asymptotic sector as ${\cal S}_{0}$,
the ``left" physical boundary condition fixes and determines the
acceptable solution on the initial sheet,
 \be
 \psi(r) = c\, \sqrt{r}\,H^{(2)}_\nu(\kappa r)\,,\ \ \ \ \
 r \in {\cal S}_0\,.
 \label{BS}
 \ee
After the $N$ counterclockwise turns of our integration path
${\cal C}^{(N)}$  around the origin this solution gets transformed
in accordance with formula 8.476.7 of ref.~\cite{Ryzhik} which
plays a key role also in some other solvable models \cite{CJT},
 \be
 H^{(2)}_\nu\left (ze^{{\rm i}m\pi}\right )=
 \frac{\sin (1+m)\pi\nu}{\sin \pi \nu}\,
 H^{(2)}_\nu(z)+
 e^{{\rm i}\pi\nu}\,
 \frac{\sin m\pi\nu}{\sin \pi \nu}\,
 H^{(1)}_\nu(z)\,.
 \label{nuit}
 \ee
Here we have to set $m=2N$. This means that the existence of a
bound state will be guaranteed whenever we satisfy the ``right"
physical boundary condition, i.e., whenever we satisfy the
elementary requirement of the absence of the unphysical component
$H^{(1)}_\nu(z)$ in the right-hand side of eq.~(\ref{nuit}).

The latter requirement is equivalent to the doublet of conditions
 \be
  2N\nu = integer\,,\ \ \ \ \ \nu \neq integer\,.
 \ee
This means that at any fixed and positive value of the energy
$E=\kappa^2$ and at any fixed winding number $N=1,2,\ldots$, our
present quantum-knot model generates the series of the bound
states at certain irregular sequence of angular momenta avoiding
some ``forbidden" values,
 \be
 \ell =\frac{M-N}{2N}\,,\ \ \ \ \
 M =  1, 2, 3, \ldots\,,\ \ \ \ \
 M \neq 2N, 4N, 6N, \ldots\,.
 \label{formu}
 \ee
These bound states exist and have the analytically continued
Hankel-function form (\ref{BS}) {\em if and only if} the {\em
kinematical input} represented by the angular momenta $\ell$ is
{\em restricted to the subset} represented by formula
(\ref{formu}).

Our construction is completed. Once we restrict our attention to
the purely kinematic model with $\gamma=0$, we can summarize that
at the odd dimensions $D=2p+1$ giving $\ell=n+p-3/2$ we may choose
any index $n$  and verify that formula (\ref{formu}) can be read
as a definition of the integer quantity $M=(2n+2p-1)\,N$ which is
not forbidden. At the even dimensions $D=2p$ we equally easily
verify that the resulting $M$ is always forbidden so that our
quantum-knot bound states do not exist at $V(r)=0$ at all.

The latter dichotomy appears reminiscent of its well-known
non-quantum real-space analogue, but the parallel is misleading
because in quantum case the freedom of employing an additional
coupling constant $\gamma$ enables us to circumvent the
restrictions. Indeed, once we select {\em any} dimension $D$,
angular-momentum index $m$, winding number $N$ and any ``allowed"
integer $M$, our spectral recipe (\ref{formu}) may simply be
re-read as an explicit definition of the knot-supporting value of
the coupling constant
 \ben
 \gamma=\left (\frac{M}{2N}
 \right )^2-
 \left (m+\frac{D-2}{2}
 \right )^2\,.
 \een
This implies that at non-vanishing $\gamma$s, the quantum knots do
exist in any dimension.

\section{Discussion \label{drzka}}


In the language of physics, our present construction and solution
of a new and fairly unusual exactly solvable quantum model of
bound states is based on the freedom of choosing the knot-shaped,
{\em complex} contours of integration ${\cal C}$. This trick is
not new \cite{BG,BB} and may be perceived as just a consequence of
the admitted loss of the observability of the coordinates in
PT-symmetric Quantum Mechanics \cite{Carl}.

From an experimentalist's point of view, the omission of the
standard assumption that the coordinate ``should be" an observable
quantity is not entirely unacceptable since the current use of the
concept of quasi-particles paved the way for similar
constructions. Related Hamiltonians could be called, in certain
sense, manifestly non-Hermitian. Still, they are currently finding
applications in nuclear physics (where they are called
quasi-Hermitian \cite{Geyer}). The loss of the reality of the
coordinates is also quite common in field theory where the similar
unusual Hamiltonians are being rather called CPT-symmetric
\cite{BBJ} or crypto-Hermitian \cite{Smilga}.

In a pragmatic phenomenological setting, the fairly unusual nature
of the new structures of spectra seems promising. At the same
time, the formalism itself is now considered fully consistent with
the standard postulates of quantum theory. In the language of
mathematics, the emergence of its innovative features may be
understood as related to non-locality, i.e., to the replacement of
the standard scalar product
 \ben
 \langle \psi\,|\,\phi\rangle = \int \psi^*(x) \phi(x) dx
 \een
by its generalized, nonlocal modifications \cite{Carl,workshops}
 \ben
 \langle \psi\,|\,\phi\rangle = \int \psi^*(x)\,
 \Theta(x,y)\,\phi(y) dx \, dy\,.
 \een
Although this leaves an overall mathematical consistency and
physical theoretical framework of Quantum Theory virtually
unchanged \cite{Carl}, a new space is being open, {\it inter
alii}, to the topology-based innovations. In principle, they might
inspire new developments of some of the older successful
applications of the formalism ranging from innovative
supersymmetric constructions \cite{susy} to cosmology
\cite{Alicos}, occasionally even leaving the domain of quantum
physics \cite{my}.


In section \ref{prva} we summarized briefly the key ingredients of
quantum theory where the ``correct" metric is assumed nontrivial,
$\Theta \neq I$. Let us now add a few comments which may have
emerged during our subsequent transition to the quantum-knot
models of section \ref{soudruzka}. Of course, our eq.~(\ref{SEr})
at $N>1$  can still be treated as compatible with the standard
postulates of quantum theory in principle. We only have to repeat
that the necessary proof of the latter compatibility statement is
nontrivial.  For each individual Hamiltonian (with the property $H
\neq H^\dagger$ with respect to the specific Dirac's definition of
the $^\dagger-$conjugation) the rigorous demonstration is
indispensable that the spectra are real and that they are discrete
and bounded from below. This demonstration represents, in fact,
the main part of our present contribution.

It is precisely the difficulty of the latter step which motivated
our present start from the dynamically trivial version (\ref{SEr})
of the BG model with vanishing $\lambda$. In the nearest future we
shall have to pay attention to the related operator $\Theta(x,y)$,
feeling inspired by the Mostafazadeh's \cite{Ali} explicit formula
 \be
 \Theta(x,y)\ \approx\
 \sum_{n=0}^{M}\,\Psi_n(x)\,s_n\,\Psi_n(y)\,,\ \ \ \ \ \ \
 M \gg 1\,
 \ee
where the normalized eigenstates $\Psi_n(x)$ of $H^\dagger$ have
to be constructed in ${\cal H}^{(original)}$ and where the real
and positive constants $s_n>1$ are, in principle, arbitrary
\cite{zno}. Of course, the letter-format of our present message
does not allow us to get too far beyond the citation of the
encouraging observation that this type of formula exhibited a
quick convergence to the exact $\Theta$ in the square-well model
where a fairly good approximation has already been obtained at $M
\approx 10$ \cite{Batal}.
%

\section{Summary \label{ruch} }

In spite of the absence of any confining force, our
Schr\"{o}dinger eq.~(\ref{SEfree}) defined along topologically
nontrivial integration paths has been shown to generate certain
bound states $\psi (r)\in {\it I\!\!\!L}^2\left ({\cal C}^{(N)}
\right ) $ at a discrete set of the centrifugal coupling $\gamma$.
We may emphasize that in such an exemplification of the more or
less standard quantum theory

\begin{itemize}

\item  the complexified coordinates are loosing their immediate
observability,

\item  arbitrary complex potentials  $V(r) \in l\!\!\!C$ are
allowed, provided only that the spectrum remains real,

\item  a  redefinition of the inner product in the Hilbert space
is required in order to return to the standard probabilistic
framework of quantum theory,

\item a challenging general open problem arises concerning the
role and tractability of
 the complex coordinate
paths with a nontrivial topological structure.

\end{itemize}

 \noindent
Of course, our present, exactly solvable  $N>1$ quantum-knot
bound-state problem would probably become purely numerical after
its immersion in virtually any external confining potential. In
this setting, even the question of survival of the reality of the
new bound-state spectra at $\lambda \neq 0$ remains open.

\section*{Acknowledgements}

Supported by the GA\v{C}R grant Nr. 202/07/1307, by the M\v{S}MT
``Doppler Institute" project Nr. LC06002 and by the NPI
Institutional Research Plan AV0Z10480505.


\section*{Figure captions}

\subsection*{Figure 1. Sample of the curve ${\cal C}^{(N)}$ with
$N=1$.}

\subsection*{Figure 2. Sample of the curve ${\cal C}^{(N)}$ with
$N=2$.}

\subsection*{Figure 3. Sample of the curve ${\cal C}^{(N)}$ with
$N=3$.}


\newpage

\begin{thebibliography}{00}

\bibitem{pra}
M. Znojil,
Phys. Rev. A 61 (2000) 066101 (quant-ph/9811088).

\bibitem{BG}
V. Buslaev and V. Grecchi, J. Phys. A: Math. Gen. 26 (1993) 5541.

\bibitem{BB}
C. M. Bender and K. Milton, Phys. Rev. D 55 (1997) R3255;

C. M. Bender and S. Boettcher, Phys. Rev. Lett. {80} (1998) 5243.

\bibitem{Batal}
A. Mostafazadeh and A. Batal, J. Phys. A: Math. Gen. 37 (2004)
11643.

\bibitem{tob}
M. Znojil, Phys. Lett. A 342 (2005) 36.

\bibitem{tobscatt}
M. Znojil, J. Phys. A: Math. Gen. 39 (2006) 13325.

\bibitem{tob2}
M. Znojil, Phys. Lett. A,
to appear (arXiv:0708.0087v1 [quant-ph] 1 Aug 2007,
doi:10.1016/j.physleta.2007.07.072).

\bibitem{ptho}
M. Znojil, Phys. Lett. A 259 (1999) 220.

\bibitem{Jones}
H. F. Jones and J. Mateo, Phys. Rev. D 73 (2006) 085002;

H. F. Jones, J. Mateo  and R. J. Rivers, Phys. Rev. D 74 (2006)
125022.

\bibitem{Geyer}
F. G. Scholtz, H. B. Geyer and F. J. W. Hahne, Ann. Phys. (NY) 213
(1992) 74.

\bibitem{Messiah}
A. Messiah, Quantum Mechanics (North Holland, Amsterdam, 1961).

\bibitem{tdep}
%
A. Mostafazadeh,  Phys. Lett. B 650 (2007) 208;

M. Znojil, Time-dependent quasi-Hermitian Hamiltonians and the
unitarity of quantum evolution, arXiv: 0710.5653 [quant-ph];

A. Mostafazadeh, Comment on ``Time-dependent quasi-Hermitian
Hamiltonians and the unitary quantum evolution'',
 arXiv: 0711.0137 [quant-ph];

M. Znojil, Reply to Comment on ``Time-dependent quasi-Hermitian
Hamiltonians and the unitary quantum evolution'',
 arXiv: 0711.0514 [quant-ph];

A. Mostafazadeh, Comment on "Reply to Comment on Time-dependent
Quasi-Hermitian Hamiltonians and the Unitary Quantum Evolution",
arXiv: 0711.1078 [quant-ph].

\bibitem{whichop}
M. Znojil, Which operator generates time evolution in Quantum
Mechanics? arXiv: 0710.0535 [quant-ph].

\bibitem{BpT}
see e.g. "branch point" in http://eom.springer.de.


\bibitem{Alvarez}
Y. Sibuya, Global Theory of Second Order Linear Differential
Equation with Polynomial Coefficient, North Holland, Amsterdam,
1975;

G. Alvarez, J. Phys. A: Math. Gen. 27 (1995) 4589.

\bibitem{Ryzhik}
I. S. Gradshteyn and I. M. Ryzhik, Tablicy integralov, summ,
ryadov i proizvedenii, Nauka, Moscow, 1971.
%

\bibitem{CJT}
F. Cannata, G. Junker and J. Trost, Phys. Lett. A 246 (1998) 219.

\bibitem{Carl}
C. M. Bender, Reports on Progress in Physics 70 (2007) 947.
%

\bibitem{BBJ}
C. M. Bender, D. C. Brody and H. F. Jones, Phys. Rev. Lett. 89
(2002) 0270401 (quant-ph/0208076).

\bibitem{Smilga}
A. V. Smilga, Cryptogauge symmetry and cryptoghosts for
crypto-Hermitian Hamiltonians, arXiv:0706.4064.

\bibitem{workshops}
H. Geyer, D. Heiss and M. Znojil, editors, J. Phys. A: Math. Gen.
39 (2006), Nr. 32 (dedicated special issue:
 pp. 9965 - 10261).
%
%

\bibitem{susy}
C. M. Bender and K. A. Milton,
%
Phys. Rev. D 57 (1998) 3595;

M. Znojil, F. Cannata, B. Bagchi and R. Roychoudhury,
Phys. Lett. B 483 (2000) 284;

A. Mostafazadeh, Nucl. Phys. B 640 (2002) 419;

M. Znojil,
J. Phys. A: Math. Gen. 35 (2002) 2341.

\bibitem{Alicos}
A. Mostafazadeh, Class. Quantum Grav. 20 (2003) 155.

\bibitem{my}
U. Guenther, F. Stefani and M. Znojil,
J. Math. Phys. 46  (2005) 063504.
%
%
%

\bibitem{Ali}
A. Mostafazadeh, J. Math. Phys. 43 (2002) 205 (math-ph/0107001)
and 2814 (math-ph/0110016).

\bibitem{zno}
M. Znojil, SIGMA 4 (2008), 001, 9 pages, arXiv: 0710.4432v3
[math-ph].


\end{thebibliography}
\end{document}